\documentclass[aps,11pt,nofootinbib,preprintnumbers,showpacs]{revtex4}

\usepackage{amssymb, amsfonts, amsmath, bm}
\usepackage[dvipdfmx]{graphicx,color}
\usepackage{ulem}

\begin{document}

\title{
A stationary and biaxisymmetric four-soliton solution in five dimensions
}
\vspace{2cm}

\author{Shinya Tomizawa${}^{1}$\footnote{tomizawasny@stf.teu.ac.jp} and Takashi Mishima${}^2$\footnote{tmishima@phys.ge.cst.nihon-u.ac.jp}}
\vspace{2cm}
\affiliation{
${}^1$ Department of Liberal Arts, Tokyo University of Technology, 5-23-22, Nishikamata, Otaku, Tokyo, 144-8535, Japan, \\
${}^2$ Laboratory of Physics, College of Science and Technology, Nihon University,
Narashinodai, Funabashi, Chiba 274-8501, Japan}

\begin{abstract} 
Using the inverse scattering method to solve the five-dimensional vacuum Einstein equations, we construct an asymptotically flat  four-soliton solution as a stationary and bi-axisymmetric solution.   
We impose certain boundary conditions on this solution so that it includes a rotating black hole whose horizon-cross section is topologically a lens space of $L(2,1)$.
The solution has nine parameters but three only is physically independent due to the constraint equations. 
The remaining  degrees of freedom correspond to the mass and two independent angular momenta of the black hole. 
We analyze a few simple cases in detail, in particular, the static case with two zero-angular momenta and the stationary case with a single non-zero angular momentum.
\end{abstract}

\pacs{04.50.+h  04.70.Bw}
\date{\today}
\maketitle

\section{Introduction}
\label{sec:1}

The studies on higher-dimensional black hole solutions to Einstein's equations have played roles in the microscopic derivation of Bekenstein-Hawking entropy~\cite{Strominger:1996sh}, and the realistic black hole production at an accelerator in the scenario of large extra dimensions~\cite{Argyres:1998qn}.  
Despite two decades of research and development in techniques of solution-generation, our understanding of higher dimensional black holes is still not enough. 
 The topology theorem for a stationary black hole generalized to five dimensions~\cite{Galloway:2005mf,Cai:2001su,Hollands:2007aj,Hollands:2010qy} states that the topology of the spatial cross section of the event horizon must be either  a sphere $S^3$, a ring $S^1\times S^2$ or lens spaces $L(p,q)$ ($p,q$ : coprime integers), if the spacetime is asymptotically flat and admits two commuting axial Killing vector fields which also commutes a stationary timelike Killing vector. 
 As for the first two topologies, the exact solutions to vacuum Einstein's equations~\cite{Tangherlini:1963bw,Myers:1986un,Emparan:2001wn,Pomeransky:2006bd} have already been found. 
In contrast, a regular vacuum black hole solution with the horizon of lens space topology has been difficult to find in spite of a few trials,   since the resultant solutions always suffer from naked singularities.

\medskip
The inverse scattering method (ISM) is known as one of the most useful tools to obtian exact solutions of Einstein equations with $D-2$ Killing isometries ($D:$ spacetime dimension). 
In this method, new solutions with the same isometries can be systematically obtained by the soliton transformation from a certain known simple solution, which is often called seed. 
In general, the direct application of the original method formulated by Belinski and Zakharov~\cite{Belinski:2001ph,Belinsky:1979mh,book exact solution} to higher dimensions yields singular solutions but  
Pomeransky modified the ISM so that it can generate regular solutions even in higher dimensions~\cite{Pomeransky:2005sj}. 
Remarkably, combined with the rod structure~\cite{Harmark:2004rm}, it has achieved a great success so far, concerned with, in particular, five-dimensional vacuum black hole solutions. 
The first example of the generation of black hole solutions by the modified ISM is the re-derivation of the five-dimensional Myers-Perry black hole solution~\cite{Pomeransky:2005sj}.  
Thereafter, the $S^2$-rotating black ring  was  re-derived~\cite{Tomizawa:2005wv} by the ISM from the Minkowski seed (this solution was first derived in Ref.~\cite{Mishima:2005id,Figueras:2005zp} independently), 
but it turned out that the generation of the $S^1$ rotating black ring  has a more delicate problem on how to choose the seed, since an facile choice of the seed always results in the generation of a singular solution. 
The suitable seed to derive the black ring with the $S^1$-rotation was first considered in~\cite{Iguchi:2006rd,Tomizawa:2006vp}. 
Subsequently, the regular black ring solution with both of $S^1$ and $S^2$ rotations was constructed by Pomerasnky and Sen'kov~\cite{Pomeransky:2006bd}.

\medskip
Using the ISM, a few authors attempted to construct asymptotically flat black lens solutions to the five-dimensional vacuum Einstein equations. 
First, Evslin~\cite{Evslin:2008gx} attempted to construct a static black lens with the lens space topology of $L(n^2+1,1)$ but  found that curvature singularities cannot be eliminated, whereas both conical and orbifold singularities can be removed.  
Subsequently, Chen and Teo~\cite{Chen:2008fa} constructed a black lens solution with the horizon topology of $L(n,1)=S^3/{\mathbb Z_n}$ by the ISM but observed that it must have either conical singularities or naked curvature singularities. 
Thus, the major obstacle in constructing a black lens solution is always suffering from naked singularities. 
However, the sudden breakthrough in this line have come from supersymmetric solutions. 
Based on the well-known framework of the construction for supersymmetric solutions in the bosonic sector of five-dimensional minimal supergravity developed by Gauntlett {\it et al.}~\cite{Gauntlett:2002nw}, 
Kundhuri and Lucietti~\cite{Kunduri:2014kja} succeeded in the derivation of the first regular exact solution of an asymptotically flat black lens with the horizon topology of $L(2,1)=S^3/{\mathbb Z}_2$. 
This solution was subsequently generalized to the more general supersymmetric black lens with the horizon topology  $L(n,1)=S^2/{\mathbb Z}_n\ (n\ge 3)$ in the same theory~\cite{Tomizawa:2016kjh}.

\medskip
The supersymmetric solutions provides us various useful information on the corresponding vacuum solutions when we do not yet know them. 
In particular,  the discovery of the supersymmetric black lens solutions~\cite{Kunduri:2014kja,Tomizawa:2016kjh} gives us a nice  guideline for understanding  the rod structure of unfound regular vacuum black lens solutions.  
From now,  for simplicity, we consider the vacuum solution with the horizon of the special lens space topology $L(2,1)$.
Figure 1 shows the rod diagram of the Kunduri-Lucietti's supersymmetric black lens in~\cite{Kunduri:2014kja}, where the horizon rod is drawn as a point because the supersymmetric black hole has a degenerate horizon. 
On the contrary, Chen and Teo in~\cite{Chen:2008fa}  considered the black lens with the rod structure displayed in FIG.2, though  this solution has conical singularities on $z\in[z_3,z_4]$\footnote{More precisely, this has two branches: one has conical singularities only on the axis, whereas the other has curvature singularities on the surface surrounding the point $z=z_4$. Here we consider the former case.}.
 The main difference between these rod diagrams except the shapes of the horizon rods lines in the signatures of the 3rd component of the rod vectors ($0,2,\pm1$) on $[z_3,z_4]$. 
 In this paper, we consider to construct the vacuum solution of a black lens with the combined rod diagram displayed in FIG.4 where the rod vector on $[z_3,z_4]$ in FIG.2 is replaced with one in FIG.1.

\begin{figure}[h]
 \begin{tabular}{cc}
 \begin{minipage}[t]{0.4\hsize}
 \centering
\includegraphics[width=8cm]{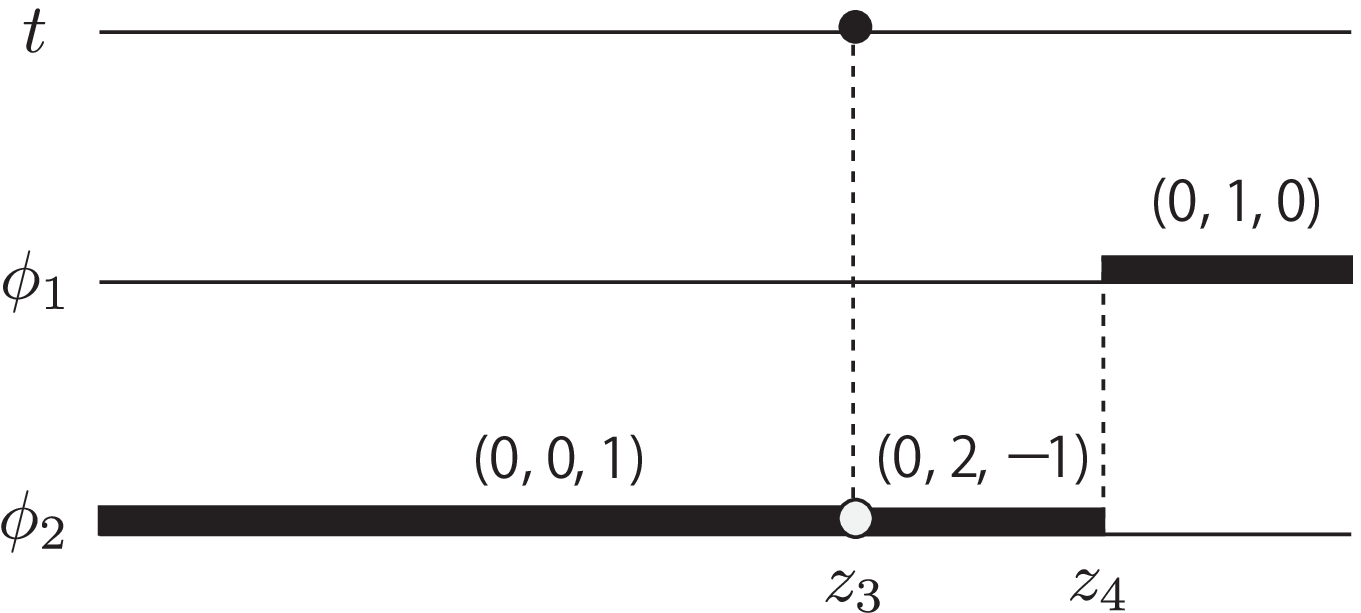}
\caption{The rod diagram of the  Kunduri-Lucietti's supersymmetric black lens with the horizon topology of $L(2,1)=S^3/{\mathbb Z}_2$.}
\label{fig:rod-KL} 
\end{minipage} &\ \ \ \ \ \ \ \ \ \ \ \ \ \ \ \ \

\begin{minipage}[t]{0.4\hsize}
 \centering
\includegraphics[width=8cm]{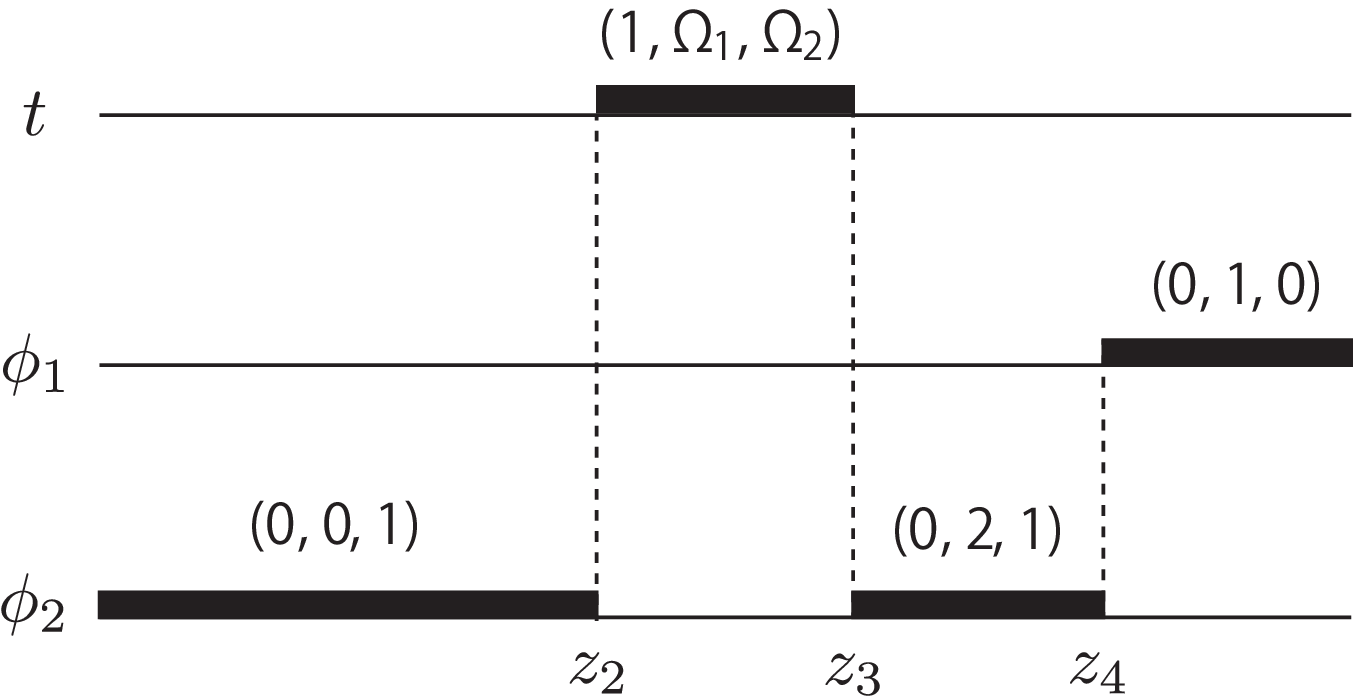}
\caption{The rod diagram of the Chen-Teo's solution with the horizon topology of $L(2,1)=S^3/{\mathbb Z}_2$.}
\label{fig:rod-CT}
\end{minipage} \\& \\
\end{tabular}
\end{figure}

\medskip

\begin{figure}[h]
 \begin{tabular}{cc}
 \begin{minipage}[t]{0.4\hsize}
 \centering
\includegraphics[width=8cm]{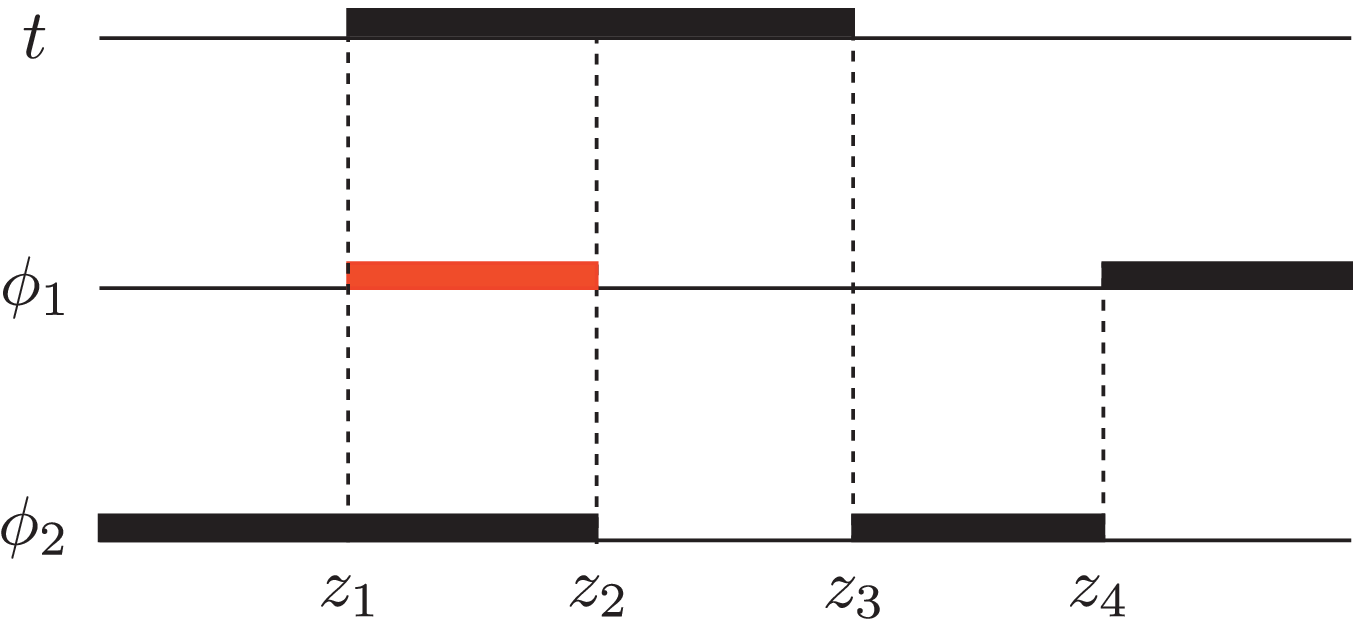}
\caption{The rod diagram of the seed solution}
\label{default}
\end{minipage} &\ \ \ \ \ \ \ \ \ \ \ \ \ \ \ \ \ 

\begin{minipage}[t]{0.4\hsize}
 \centering
\includegraphics[width=8cm]{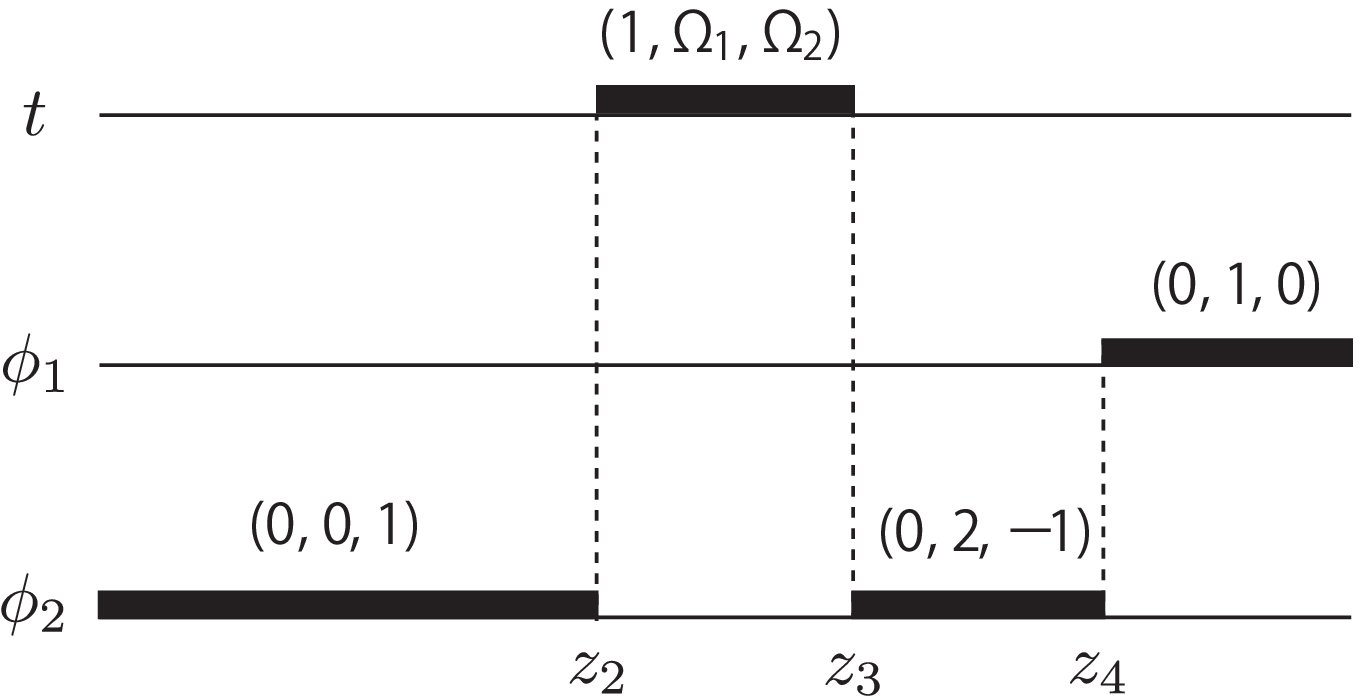}
\caption{The rod diagram of the obtained four-soliton solution.}
\label{fig:rod-TM}
\end{minipage} \\& \\
\end{tabular}
\end{figure}

 \medskip
To this end, using the Pomeransky's ISM for the five-dimensional vacuum Einstein equation, we construct a four-soliton solution by regarding the singular solution with a diagonal metric as a seed (see FIG.3 on the rod diagram), which is the same as one used for the construction of the black ring solution. 
In order to obtain a black lens solution of physical interest, we impose suitable boundary conditions at infinity, on the horizon, on a symmetry of axis as follows: 
 (i)  Infinity must be such that the spacetime is asymptotically flat. 
(ii) The horizon corresponds to a smooth null surface, whose spatial cross section has a topology of the lens space $L(2,1)=S^3/{\mathbb Z}_2$.
 (iii) On the axis, there appear no curvature singularities, no conical singularities,  no Dirac-Misner strings, and besides, orbifold singularities at isolated points must be eliminated. There indeed exist the parameter regions where all these boundary conditions are satisfied, however, 
 it seems to be considerably hard to deal with the four-soliton itself owing to the rather complex metric and five constraints on the parameters. 
 For this reason, when we study the physical properties of the solution, we restrict ourselves to a few simple cases,  a case with a single angular momentum and a static case.
As shown later, as for the  the black lens with a single angular momentum, there exist unavoidable naked CTCs surrounding the nut $z=z_4$ outside the horizon, even if all of (conical, curvature, and orbifold) singularities and Dirac-Misner strings can be removed at these boundaries. 
For the static case,  there are two branches, though there necessarily exist conical singularities on $[z_3,z_4]$ only. 
One has CTCs region around the point $z=z_4$, but the other does not.

\medskip
We organize the present paper as follows. 
In the following section~\ref{sec:solution}, by using the Pomeransky's ISM, we present the four-soliton solution in five dimensions which admits three commuting Killing vectors,  stationary and bi-axially symmetric Killing vectors. 
The solution contains many parameters. 
In the general choice of these parameters, the solution cannot be necessarily asymptotically flat and regular, and the lens space topology of the horizon is not guaranteed.
Therefore, in section~\ref{sec:boundary}, we impose on the parameters the boundary conditions under which the spacetime 
is asymptotically flat, neither (curvature, conical, and orbifold) singularities nor Dirac-Misner strings exist, at least, on the axis and horizon.  
It is shown that the boundary conditions  finally reduce the number of the independent parameters to three. 
In Section~\ref{sec:simple case}, we  analyzes a few simple cases,  the case with a single angular momentum and a static case. 
In particular, for the former case, we will discuss the phase diagram and the existence of CTCs.    
 In the final section~\ref{sec:summary}, we devote ourselves to the summary and discussion on our results.

\section{Black lens solutions}
\label{sec:solution}
First, let us start from  the construction of the seed solution.
We consider a five-dimensional, stationary and bi-axisymmetric spacetime whose has three commuting commuting Killing vectors, a stationary Killing vector $\partial/\partial t$, and two axisymmetric Killing vectors  $\partial/\partial \phi_1$, $\partial/\partial \phi_2$.
The diagonal metric of the five-dimensional vacuum solution whose rod diagram is given by FIG.3 can be written as
\begin{eqnarray*}
ds^2=-\frac{\mu_1}{\mu_3}dt^2+\frac{\mu_2\mu_4}{\mu_1}d\phi_1^2+\frac{\rho^2\mu_3}{\mu_2\mu_4}d\phi_2^2+k^2\frac{\mu_3R_{11}R_{22}R_{33}R_{44}R_{24}^2}{\mu_2\mu_4R_{12}R_{13}R_{14}R_{23}R_{34}}(d\rho^2+dz^2), 
\end{eqnarray*}
where $\mu_i$ and $R_{ij}\  (i,j=1,\ldots,4)$ are defined, respectively,  by
\begin{eqnarray*}
\mu_i=\sqrt{\rho^2+(z-z_i)^2}-(z-z_i),\quad R_{ij}=\frac{\mu_i\mu_j}{\rho^2+\mu_i\mu_j}.
\end{eqnarray*}
$z_{i}$ are constants,  and we assume $z_1<z_2<z_3<z_4$. 
As seen later, $k$ is the integration constant which is determined from the requirements of the absence from conical singularities at infinity. 
Note that this metric is the exactly same as that of the seed solution used for the derivation of the rotating black ring by the ISM.
As shown later, although there exist naked curvature singularities on the rod $\rho=0, z\in [z_1,z_2]$ which has a negative density,  
they have disappeared under some appropriate boundary conditions after the four-soliton transformation. 
Now, we briefly explain how we have obtained the four-soliton solution, following the Pomeransky's procedure:

\begin{itemize}
\item First, let us remove trivial solitons from the four points $z=z_1,z_2,z_3$, and $z_4$ with the BZ vectors $(0,1,0),\ (0,0,1),\ (1,0,0)$ and $(0,0,1)$, respectively.
\begin{itemize}
\item After the solitons are removed, the obtained metric  is written as
\begin{eqnarray}
\tilde g_0&=&g_0 \cdot {\rm diag}\left( -\frac{\rho^2}{\tilde \mu_3^2},-\frac{\rho^2}{\tilde \mu_1^2},\left(-\frac{\rho^2}{\tilde \mu_2^2}\right) \left(-\frac{\rho^2}{\tilde \mu_4^2}\right)\right)\\
             &=& {\rm diag}\left(\frac{\rho^2}{\tilde \mu_1\tilde\mu_3},\frac{\rho^4}{\tilde\mu_1\tilde\mu_2\tilde\mu_4},-\frac{\rho^4}{\tilde\mu_4\tilde\mu_3\tilde\mu_4} \right),
\end{eqnarray}
where $\tilde \mu_i=-\rho^2/\mu_i=-\sqrt{\rho^2+(z-z_i)^2}-(z-z_i)$.
\item  Performing the formal replacement of $\tilde\mu_i\to \tilde\mu_i-\lambda\ (i=1,\ldots,4)$, $\rho^2 \to \rho^2-2z\lambda -\lambda^2$ in the matrix $\tilde g_0$ ($\lambda$ is a so-called spectrum parameter), we can obtain the generating matrix $\Psi_0[\lambda,\rho,z]$ corresponding to $\tilde g_0$:
\begin{eqnarray}
\Psi_0[\lambda,\rho,z]
&=&{\rm diag}\biggl(
\frac{\rho^2-2z\lambda-\lambda^2}{(\tilde \mu_1-\lambda)(\tilde \mu_3-\lambda)},
\frac{(\rho^2-2z\lambda-\lambda^2)^2}{(\tilde\mu_1-\lambda)(\tilde\mu_2-\lambda)(\tilde \mu_4-\lambda)},\notag\\
&&\hspace{5cm} -\frac{(\rho^2-2z\lambda-\lambda^2)^2}{(\tilde\mu_2-\lambda)(\tilde\mu_3-\lambda)(\tilde \mu_4-\lambda)}
\biggr).
\end{eqnarray}
\end{itemize}

\item Next, let us add back the nontrivial solitons at $z=z_1,z_2,z_3$ and $z_4$ with the BZ vectors $m_{0a}^{(1)}=(C_1,1,0),m_{0a}^{(2)}=(0,C_2,1),m_{0a}^{(3)}=(1,0,C_3)$ and $m_{0a}^{(4)}=(0,C_4,1)$, respectively, where the constants $C_1,C_2,C_3$ and $C_4$ are often called BZ parameters.

Then,  using the BZ vectors and the inverse matrix of the generating matrix whose spectrum parameter $\lambda$ is substituted into $\mu_k$ , we can construct the three-dimensional vectors $m_a{}^{(k)}$,
\begin{eqnarray}
m_a^{(k)}=m_{0b}^{(k)}[\Psi_0(\mu_k,\rho,z)]_{ba}.
\end{eqnarray}
Thus, we obtain the metric of the four-soliton solution as
\begin{eqnarray}
g_{ab}=(\tilde g_0)_{ab}-\sum_{k,l=1}^4\frac{(g_0)_{ac}m_c{}^{(k)}(\Gamma^{-1})_{kl}m_d{}^{(l)}(g_0)_{bd}}{ \mu_k\mu_l},
\end{eqnarray}
\begin{eqnarray}
\Gamma_{kl}=\frac{m_a{}^{(k)}(\tilde g_0)_{ab}m_b{}^{(l)}}{\rho^2+\mu_k\mu_l}. \label{eq:Gamma}
\end{eqnarray}
 Note that in the original BZ's procedure, $g_{ab}$  does not satisfy the normalization condition ${\rm det} (g_{ab})=-\rho^2$ (in the final step one must normalize $g_{ab}$),  whereas in the Pomeransky's procedure, $g_{ab}$ automatically satisfies the condition without such a normalization process. 
The two-dimensional  conformal factor $f:=g_{\rho\rho}(=g_{zz})$ for the new solution is obtained from the factor $f_0$ for the seed  as
\begin{eqnarray}
f=f_0\frac{{\rm det}\ \Gamma}{{\rm det}\ \Gamma_0},
\end{eqnarray}
where the matrix $\Gamma_0=(\Gamma_{0kl})$ is obtained by putting  $C_i=0$ in Eq.(\ref{eq:Gamma}).
\end{itemize}

\section{Boundary conditions}\label{sec:boundary}
The four-soliton solution that we have obtained in the previous section has nine parameters $(z_i,C_i,k)$ but 
the general choice of these parameters cannot guarantees the horizon topology of a lens space,  regularity  and even asymptotic flatness.
 In order that the four-soliton solution describes a physically interesting solution, we need impose suitable boundary conditions at infinity, on the horizon, and on a symmetry of axis : %
(i) The spacetime is asymptotically flat at infinity. 
(ii) The spacetime has a smooth horizon whose spatial topology is the lens space $L(2,1)=S^3/{\mathbb Z}_2$.
 (iii) The spacetime has  no curvature singularities, no conical singularities,  no Dirac-Misner strings, and besides, no orbifold singularities at isolated points on the axis. 
In what follows, using the rod structure~\cite{Harmark:2004rm}, we  show that there indeed exist the parameter regions where all these boundary conditions are satisfied.

\subsection{rod diagram}

To ensure that the spacetime is asymptotically flat,  the two semi-infinite rods $(-\infty,z_1]$ and $[z_4,\infty)$ must have the rod vectors $(0,0,1)$ and $(0,1,0)$, respectively.  At a glance, this seems not to be satisfied for the obtained metric, but we can confirm that it is automatically satisfied under the global rotation $g\to A^TgA $, where $A$ is a $3\times3$ matrix that satisfies the condition ${\rm det} \ A=1$ and is written as
\begin{eqnarray}
A&=&\zeta \left(
    \begin{array}{ccc}
      \zeta^{-1} & \frac{C_3 (-C_2 z_{12} z_{34}-C_4 z_{14} z_{23})}{C_1 C_3 z_1 (C_2 z_{34}+C_4 z_{23})+z_3 z_{24}}& \frac{2 C_1 z_1 z_3 (C_2-C_4) z_{13}}{C_1 C_3 z_1 (C_2 z_{34}+C_4 z_{23})+z_3 z_{24}} \\
      0 & 1 & -\frac{C_1 C_2 C_3 C_4 z_1 z_{24}+z_3 (C_2 z_{21}+C_4 z_{14})}{C_1 C_3 z_1 (C_2 z_{34}+C_4 z_{23})+z_3 z_{24}} \\
      0 &  -\frac{C_1 C_2 C_3 C_4 z_1 z_{24}+z_3 (C_2 z_{21}+C_4 z_{14})}{C_1 C_3 z_1 (C_2 z_{34}+C_4 z_{23})+z_3 z_{24}}
 & 1
    \end{array}
  \right),
\end{eqnarray}
where $z_{ij}:=z_i-z_j$, and $\zeta$, which is determined from ${\rm det} A=1$, can be written as
\begin{eqnarray}
\zeta&=&-\left[C_1 C_3 z_1 (C_2 z_{43}+C_4 z_{32})+z_3 z_{42}\right]\notag\\
&\times&\biggl[ z_3(C_2z_{21}-C_4z_{41}-z_{42})-C_1C_3z_1(C_2C_4z_{42}+C_2z_{43}+C_4z_{32}) \biggr]^{-\frac{1}{2}}\notag\\
&\times&\biggl[ z_3(C_2z_{21}-C_4z_{41}+z_{42})-C_1C_3z_1(C_2C_4z_{42}-C_2z_{43}-C_4z_{32}) \biggr]^{-\frac{1}{2}}.
\end{eqnarray}

\begin{itemize}
\item The semi-infinite rod $z\in(-\infty,z_1]$ has the rod vector 
\begin{eqnarray}
v_1&=&\left( \frac{2z_1z_3z_{31}C_1(C_2-C_4)}{z_1C_1C_3( z_{32}C_4+z_{43}C_2 )+z_3z_{42}},\frac{-z_1z_{42}C_1C_2C_3C_4+z_3(z_{21}C_2-z_{41}C_4)}{z_1C_1C_3( z_{32}C_4+z_{43}C_2 )+z_3z_{42}},1\right).\notag
\end{eqnarray}
After the global rotation, $v_1' =A^{-1}v_1$ becomes proportional to $(0,0,1)$. 
Moreover, the condition for conical singularities not to exist on the rod is given by
\begin{eqnarray}
m_{1}:=\lim_{\rho\to 0}\sqrt{\frac{\rho^2f}{g'_{ij}(v'_{1})^i(v'_{1})^j}}=1,
\end{eqnarray}
which determine $k$ as
\begin{eqnarray}
k^2&=&-\frac{z_3^2 z_{42}^2}{ z_3(C_2z_{21}-C_4z_{41}-z_{42})-C_1C_3z_1(C_2C_4z_{42}+C_2z_{43}+C_4z_{32}) }\notag\\
&\times& \frac{1}{z_3(C_2z_{21}-C_4z_{41}+z_{42})-C_1C_3z_1(C_2C_4z_{42}-C_2z_{43}-C_4z_{32}) }. \label{eq:k} 
\end{eqnarray}
\item The semi-infinite rod $[z_4,\infty)$ has the rod vector 
\begin{eqnarray}
v_4&=&\left(\frac{C_3 (C_2 z_{21} z_{43}+C_4 z_{41} z_{32})}{C_1 C_3 z_1 (C_2 z_{43}+C_4 z_{32})+z_3 z_{42}},1,\frac{-C_1 C_2 C_3 C_4 z_1 z_{42}+z_3 (C_2 z_{21}-C_4 z_{41})}{C_1 C_3 z_1 (C_2 z_{43}+C_4 z_{32})+z_3 z_{42}}\right).
\end{eqnarray}
After the global rotation, $v_{4}' =A^{-1}v_{4}$ becomes proportional to $(0,1,0)$. 
The condition (\ref{eq:k}) automatically guarantees the absence from  the conical singularities on the axis rod. 
In what follows, we define $(\phi'_1,\phi'_2)$ by $\partial/\partial\phi'_1:=v_{4}'$ and  $\partial/\partial\phi'_2:=v_{1}'$.

\item The  finite rod $[z_1,z_2]$ has the rod vector 
\begin{eqnarray}
v_{12}&=&\biggl(\frac{2 z_3 (C_2-C_4) z_{21} z_{31} z_{41}}{C_3 z_{21} z_{41} (C_2 z_{43}+C_4 z_{32})+2 C_1 z_1 z_3 z_{31} z_{42}},\notag\\
&&\frac{C_2 z_{21} (2 C_1 z_1 z_3 z_{31}-C_3 C_4 z_{41} z_{42})-2 C_1 C_4 z_1 z_3 z_{31} z_{41}}{C_3 z_{21} z_{41} (C_2 z_{43}+C_4 z_{32})+2 C_1 z_1 z_3 z_{31} z_{42}},1\biggr) .
\end{eqnarray}
To eliminate the naked curvature singularities on $[z_1,z_2]$ which indeed exists for the seed solution, 
we require that the finite rod $[z_1,z_2]$ should be parallel to the the semi-infinite rod $(-\infty,z_1]$. 
It can be easily shown that this can be accomplished if we impose the condition 
\begin{eqnarray}
C_1=\pm\sqrt{\frac{z_{21}z_{41}}{2z_1^2z_{31}}}. \label{eq:condi1}
\end{eqnarray}
It turns out that after the global rotation, $v_{12}' =A^{-1}v_{12}$ becomes proportional to $(0,0,1)$.

\item The timelike finite rod $[z_2,z_3]$ has the following rod vector 
\begin{eqnarray}
v_{23}&=&\biggl(1,
\frac{-2 C_1 z_1z_3 z_{31}  (z_{32}+C_2 C_4 z_{41})+C_3 C_4 z_{32} z_{42} z_{41}}{-2z_3z_{31}z_{41}(C_2C_4z_{21}+z_{32})},\notag\\
&&\frac{2 C_1 C_2 z_1 z_3z_{31} z_{42} +C_3 z_{41} z_{32} (C_2 C_4z_{21}-z_{43})}{-2z_3z_{31}z_{41}(C_2C_4z_{21}+z_{32})}\biggr).
\end{eqnarray}
After the global rotation, $v_{23}' =A^{-1}v_{23}$ becomes proportional to $(1,\Omega_1,\Omega_2)$, where $\Omega_1$ and $\Omega_2$ are the angular velocities along $\partial_{\phi_1'}$ and $\partial_{\phi_2'}$, respectively, and are given by

\begin{eqnarray}
\Omega_1&=&\frac{\tilde\Omega_1}{D_1},\\
\Omega_2&=&\frac{\tilde\Omega_2}{D_2},
\end{eqnarray}
where
\begin{eqnarray}
\tilde \Omega_1&=& -4 C_1 z_1 z_3^3 z_{31}^2 z_{42}+C_3^3 C_4^2 z_{21}  z_{32}z_{41}^2 (C_4 z_{32}+C_2 z_{43})\notag\\
&&+2 C_1 C_3^2 C_4 z_1z_3 z_{31} z_{41}  [C_2 (z_{32}-z_{21}) z_{43}+C_4 z_{32} (-z_{21}+z_{32}+z_{42})]\notag\\
&&+2 C_3z_3^2 z_{31} z_{41}  [-C_2 z_{21} z_{43}+C_4 (z_{32}^2+z_{42}^2-z_{43}^2+z_1z_{42}+z_2z_{21}-z_3z_{41})],
\end{eqnarray}
\begin{eqnarray}
\tilde \Omega_2&=&-4 C_1 z_1 z_3^3z_{31}^2 z_{42}  [-C_2^2 C_4 z_{21} z_{41}+C_4 z_{41} z_{32}
+C_2 (C_4^2 z_{14}^2+z_1z_{32}+z_2z_{43}-z_4z_{42})]\notag\\
&&+2 C_3 z_3^2z_{31} z_{41}  \bigg[C_2^3 C_4 z_{21}^2 z_{41} z_{43}+C_2 C_4 z_{21} z_{32} 
(-C_4^2 z_{41}^2-z_{21}z_{43}+z_{32}z_{41})\notag\\
&&-z_{32} \{z_{42}^2 z_{43}+C_4^2 z_{41} (z_{21}z_{32}-z_{42}^2)\}\notag \\
&&-C_2^2 z_{21}\biggl\{- z_{43} (z_{21}z_{32}+2z_{42}^2)-C_4^2 z_{41} (z_{21}z_{32}-z_{41}z_{43}-z_{42}^2)\biggr\}\biggr]\notag\\
&&-C_3^3 z_{21} z_{41} z_{32} \biggl[C_4^2 z_{32}^2 z_{41} z_{43}-C_2^3 C_4 z_{21} z_{41} z_{43}^2-C_2 C_4 z_{41} z_{32} \{-2 z_{43}^2+C_4^2 (z_{21}z_{32}+z_{42}^2)\}\notag\\
&&+C_2^2  z_{41}z_{43}\{ z_{43}^2-C_4^2 (z_{21}z_{32}+z_{42}^2)\}\biggr]\notag\\
&&-2 C_1 C_3^2 z_1z_3  z_{31} z_{41} z_{42} \biggl[-C_4^3 z_{41} (z_{32}^2+C_2^2 z_{21} z_{23})+2 C_4 (z_{32}^2-C_2^2 z_{21} z_{32}) z_{43}\notag\\
&&-C_2 C_4^2 \{-C_2^2 z_{21} z_{41} z_{43}+z_{32}  z_{42}(z_{41}+z_{43})\}+C_2 z_{43} (2 z_{32} z_{43}-C_2^2 z_{21} z_{43})\biggr],
\end{eqnarray}
\begin{eqnarray}
\zeta D_1&=&\zeta D_2(z_{42}(C_2^2z_{21}+z_{32}))^{-1} \notag\\
&=&-2 C_1 C_3^3 C_4^2 z_1 z_{31} z_{41} z_{32} (C_4 z_{41} z_{32}+C_2 z_{21} z_{43})\notag \\
&&+2 C_3^2 C_4 z_3 z_{31} z_{41} [-C_4 z_{41} z_{32}^2+C_2 z_{21} (C_4^2 z_{41} z_{42}+z_{32}^2-z_{42}^2)] \notag\\
&&-4  z_3^3z_{31}^2 z_{41} (-C_4^2 z_{41}+C_2 C_4 z_{21}+z_{42}) \notag\\
&&-4 C_1 C_3 z_1 z_3^2z_{31}^2  [C_2 z_{42} z_{43}-C_4^3 z_{41} z_{41}+C_4 z_{41} z_{42}-C_2 C_4^2 z_{41} (-z_{21}+z_{42})] .
\end{eqnarray}

\item The finite rod $[z_3,z_4]$ has the rod vector
\begin{eqnarray}
v_{34}&=&\biggl(\frac{-C_3 z_{41} z_{43} (C_2 C_4 z_{21}+z_{32})}{C_2 (C_1 C_3 z_1 z_{42} z_{43}-C_4 z_3 z_{21} z_{41})+z_3 z_{41} z_{43}},\notag\\
&&-\frac{C_1 C_3 z_1 z_{34} (C_2 C_4 z_{41}+z_{32})+C_4 z_3 z_{41} z_{42}}{C_2 (C_1 C_3 z_1 z_{42} z_{43}-C_4 z_3 z_{21} z_{41})+z_3 z_{41} z_{43}},1\biggr).
\end{eqnarray}
After the global rotation, the rod vector $v_{34}'=A^{-1}v_{34}$ is proportional to $(0,2,-1)$ if the constants $(C_i,z_i)$ satisfy  
\begin{eqnarray}
C_4&=&-\frac{C_1C_2z_1(2z_3^2z_{31}+C_3^2z_{32}z_{43})+C_2^2C_3z_3z_{21}z_{42}+C_3z_3z_{32}z_{42}}{C_1z_1(C_3z_{32}^2+C_2^2C_3^2z_{21}z_{42}-2z_3^2z_{31})}, \label{eq:condi-C_4}
\end{eqnarray}
and 
\begin{eqnarray}
&&2z_{42} \left(-C_4^2 z_{41}+z_{43}\right) (C_1 C_2 C_3 z_1+z_3) (C_1C_2C_3z_{43}z_1+z_{41}z_3)\notag\\
&&-C_1^2 C_3^2 z_1^2 z_{34} \left\{C_2^2 C_4 (z_1 z_{43}-z_2 z_{42}+z_{32} z_4)-C_2 z_{23} \left(C_4^2 z_{14}+z_{34}\right)-C_4 z_{23}^2\right\}\notag\\
&&+C_1 C_3 z_1 z_{24} z_3 \left(C_2^2 z_{12}+z_{23}\right) \left(C_4^2 z_{14}+z_{34}\right)+z_{14} z_3^2 \left\{-C_2^2 C_4 z_{12}^2+C_2 z_{12} (C_4^2 z_{14}+z_{34}\right)\notag \\
&&+C_4 (z_1 z_{43}-z_2 z_{42}+z_{32} z_4)=0.\label{eq:condi-lens}
\end{eqnarray}
In fact, under these conditions, $v'_{34}=(v'_{34t},v'_{34\phi_1'},-1)$ is written as
\begin{eqnarray}
v'_{34t}&=&\biggl[-2 C_1^2 C_2 C_3 z_1^2 z_{31} z_3 (C_2-C_4)-C_1 z_1 \{C_3^2 \left(C_2^2 C_4 z_{21} z_{42}+C_2 z_{32} z_{43}+C_4 z_{32}^2\right)\notag\\
&-&2 z_3^2z_{31} (C_2-C_4) \}+C_3 z_3 \left(C_2^2 z_{21}^2-C_2 C_4 z_{21} z_{41}-z_{32} z_{42}\right)\notag \biggr]\zeta\notag \\
&\times&[(C_1 C_2 C_3 z_1+z_3) \{-C_1 C_3 z_1 (C_2 z_{43}+C_4z_{32})-z_3 z_{42}\}]^{-1} \\
&=&0,
\end{eqnarray}
and 
\begin{eqnarray}
v'_{34 \phi_1'}&=&\biggl[C_1^2 C_3^2 z_1^2 z_{34} \left\{C_2^2 C_4 (z_1 z_{43}-z_2 z_{42}+z_{32} z_4)-C_2 z_{32} \left(C_4^2 z_{41}+z_{43}\right)-C_4 z_{32}^2\right\}\notag\\
&+&C_1 C_3 z_1 z_{42} z_3 \left(C_2^2 z_{21}+z_{32}\right) \left(C_4^2 z_{41}+z_{43}\right)\notag\\
&+&z_{41} z_3^2 \left\{-C_2^2 C_4 z_{21}^2+C_2 z_{21} (C_4^2 z_{41}+z_{43}\right)+C_4 (z_1 z_{43}-z_2 z_{42}+z_{32} z_4)\}\biggr]\notag\\
&\times& [z_{42} \left(-C_4^2 z_{41}+z_{43}\right) (C_1 C_2 C_3 z_1+z_3) (C_1C_2C_3z_{43}z_1+z_{41}z_3)]^{-1}\notag \\
&=&2.
\end{eqnarray}

Moreover, the conical singularities are free on $z\in[z_3,z_4]$ if the constants satisfy 
\begin{eqnarray}
m_{34}:=\lim_{\rho\to 0}\sqrt{\frac{\rho^2f}{g'_{ij}(v'_{34})^i(v'_{34})^j}}=1,
\end{eqnarray}
which gives
\begin{eqnarray}
m_{34}^2&=&\frac{z_{42}^2 \left(C_4^2 z_{41}-z_{43}\right)^2 (C_1 C_2 C_3 z_1+z_3)^2(C_1C_2C_3z_{43}z_1+z_{41}z_3)^2 }
{z_{41} z_{43}\biggl[ z_3(C_2z_{21}-C_4z_{41}-z_{42})-C_1C_3z_1(C_2C_4z_{42}+C_2z_{43}+C_4z_{32}) \biggr]^2}\notag \\
&\times& \frac{1} {\biggl[ z_3(C_2z_{21}-C_4z_{41}+z_{42})-C_1C_3z_1(C_2C_4z_{42}-C_2z_{43}-C_4z_{32}) \biggr]^{2}}\notag\\
&=&1.\label{eq:condi-conical}
\end{eqnarray}
\end{itemize}

\subsection{Summary}
The physical requirements of asymptotic flatness, regularity on the rods and horizon topology of the lens space $L(2,1)$ impose the conditions (\ref{eq:k}), (\ref{eq:condi1}), (\ref{eq:condi-C_4}), (\ref{eq:condi-lens}) and (\ref{eq:condi-conical}) on the parameters. 
In combination with the gauge degree of freedom $z\to z+\alpha$, these reduce the independent parameters from nine to three. These correspond to physical degree of freedom, mass and two angular momenta.
The ADM mass and two ADM angular momenta are given by, respectively, 
\begin{eqnarray}
M=\frac{3\pi m}{8D},\quad J_1=\frac{\pi j_1}{4D},
\end{eqnarray}
\begin{eqnarray}
J_2=\frac{\pi }{4}\frac{2C_3(z_{32}+C_2^2z_{21})(2z_3^2z_{31}+C_3^2z_{32}z_{43})z_{42}z_{32}\zeta_2}{-C_3^2z_{32}^2+2z_3^2z_{31}-C_2^2C_3^2z_{21}z_{42}},
\end{eqnarray}

where 
\begin{eqnarray}
m&=&2z_{31}z_{42}[C_2^2z_{21}\{2z_3^2z_{31}+C_3^2(C_4^2z_{41}z_{42}-z_{43}^2)  \}+C_3^2C_4^2z_{32}^2z_{41}+2z_{31}z_3^2(z_{42}-C_4^2z_{41})   ],\\
j_1&=& 2z_{42}[2C_2^2C_3C_4z_{21}z_{31}z_{41}z_3^2(z_{42}^2+z_{32}^2-2C_4^2z_{41}z_{42})\nonumber\\
&&+C_3C_4z_{32}z_{41}\{  C_3^2C_4^2z_{32}^3z_{41}+2z_{31}z_3^2(-C_4^2(z_{42}+z_{32})+z_{42}^2) \}+C_2C_3^3C_4^2z_{41}z_{43} \nonumber\\
&&+2C_2^3C_3z_{21}^2z_{31}z_3^2\{2C_4^2z_{41}z_{42}-z_{43}(z_{42}+z_{43})\}  -2C_2C_3z_{21}z_{31} z_3^2\{  -C_4^2(z_{32}^2+z_{42}^2)+z_{42}^2z_{43}    \} \nonumber\\
&&+C_2^2C_3^3C_4z_{21}z_{32}z_{41}z_{43}(C_4^2z_{32}+z_{43})+C_2^3C_3^3z_{21}^2z_{43}^2(C_4^2z_{32}+z_{43})]\nonumber\\
&&+2C_1z_{42}[-4C_2^2z_1z_3^3z_{21}z_{31}^2(C_2C_4z_{21}-2C_4^2z_{41}+z_{42}+z_{43})\notag\\
&&-4C_2C_4z_1z_3^3z_{31}^2( C_4^2z_{41}^2+z_{21}z_{32}-z_{41}z_{43} ) \nonumber\\
&& -2z_1z_3z_{31}\{  C_3^2C_4^2z_{32}^3z_{41}+2z_3^2z_{31}(-C_4^2z_{41}(z_{42}+z_{32}))+z_{42}^2         \} \nonumber\\
&& -2C_2^3C_3^2C_4z_1z_3z_{21}z_{31}\{  C_4^2z_{41}z_{42}^2+z_{43}(z_{32}z_{21}-z_{42}^2)           \} \nonumber  \\
&&+2C_2^2C_3^2z_1z_3z_{21}z_{31}\{  -C_4^2z_{41}(z_{32}^2+z_{43}^2)+z_{43}^2        \}  \nonumber\\
&&+2C_2C_3^2C_4z_1z_3z_{31}z_{32}\{  -z_{43}(2z_{42}^2-z_1z_{43}+z_2z_{41}-z_3z_{21}+z_4z_{32})\notag\\
&& -C_4^2z_{41}(z_{42}^2+z_{41}z_{43})     \}] ,\\
D&=&C_3^2z_{21}z_{41}\{ (C_2z_{43}+C_3z_{32})^2-C_2^2C_4^2z_{42}^2  \}   -2z_{31}z_3^2 \{ (C_2z_{21}+C_4z_{41})^2-z_{42}^2  \}\nonumber \\
&&-4C_1C_3z_{31}z_{42}\{ C_2C_4(-C_2z_{21}+C_4z_{41})-C_4z_{32}-C_2z_{43}  \} .
\end{eqnarray}


\section{Limits to simple solution}\label{sec:simple case}

\subsection{A black lens with a single angular momentum} \label{sec:single}
For simplicity, we analyze a rotating black lens with only a single angular momentum, where note that  ``a single angular momentum" does not mean  ``a single angular velocity".  
In fact, the solution with a single angular momentum has two non-vanishing angular velocities.
One of such solutions can be obtained by taking the three-soliton limit of $C_3=0$, and one finds that the conditions (\ref{eq:condi-C_4}), (\ref{eq:condi-lens}) and (\ref{eq:condi-conical}) reduce, respectively, to 
\begin{eqnarray}
&&C_4= C_2, \\
&&z_{21}C_2^3+2z_{41}C_2^2+z_{32}C_2-2z_{43}=0, \label{eq:condi3-2}\\
&&\frac{z_{41}(z_{41}C_2^2-z_{43})^2}{(1-C_2^2)^2z_{43}z_{42}^2}=1.\label{eq:condi4-2}
\end{eqnarray}

It should be note that this solution is different from the Chen-Teo's solution with a single angular momentum in Ref.\cite{Chen:2008fa}, since how to add back non-trivial solitons differs in what follows.
They removed a trivial soliton with $(0,0,1)$ from $z=z_2$ and then added the non-trivial soliton with $(C_2,0,1)$ at  $z=z_2$, whereas after we remove the trivial soliton with $(0,0,1)$ from $z=z_2$, we add back the non-trivial soliton with  $(0,C_2,1)$ at $z=z_2$.

\subsubsection{C-metric representation}
For the present purpose, it is more convenient to use a so-called $C$-metric coordinates $(x,y)$ rather than the canonical coordinates $(\rho,z)$. The relation between these coordinates is given by
\begin{eqnarray}
\rho&=&\frac{2\kappa^2\sqrt{-G(x)G(y)}}{(x-y)^2},\\
z&=&\frac{\kappa^2(1-xy)\{2+\mu(x+y)\}}{(x-y)^2},
\end{eqnarray}
where
\begin{eqnarray}
G(\xi):=(1-\xi^2)(1+\mu \xi),
\end{eqnarray}
and  $\mu$ and $\kappa$ are constants, which satisfy the inequalities
\begin{eqnarray}
0\le \mu<1,\quad \kappa>0.
\end{eqnarray}
To fix the gauge freedom $z\to z+\alpha$, let the turning points $(z_1,z_2,z_3,z_4)$ be
\begin{eqnarray}
z_1=c\kappa^2,\quad z_2=-\mu\kappa^2,\quad z_3=\mu\kappa^2,\quad z_4=\kappa^2.
\end{eqnarray}
Then, we can write $(\mu_2,\mu_3,\mu_4)$ in the simple forms without the square root as
\begin{eqnarray}
\mu_2&=&-\frac{2\kappa^2(1-x)(1+y)(1+\mu y)}{(x-y)^2},\\
\mu_3&=&-\frac{2\kappa^2(1-x)(1+y)(1+\mu x )}{(x-y)^2},\\
\mu_4&=&-\frac{2\kappa^2(1-y^2)(1+\mu x )}{(x-y)^2}.
\end{eqnarray}
Before performing the global rotation mentioned previously, the metric of the rotating black lens with the horizon topology of $L(2,1)$  takes the following form

\begin{eqnarray}
ds^2&=&-\frac{H(x,y)}{H(y,x)}\left[ dt+\omega_1d\phi_1+\omega_2\phi_2\right]^2\notag\\
&&+\frac{F(x,y)}{(x-y)^2H(x,y)}d\phi_1^2-\frac{F(y,x)}{(x-y)^2H(x,y)}d\phi_2^2+2\frac{J(x,y)}{(x-y)^2H(x,y)}d\phi_1d\phi_2\notag\\
&&+\frac{\kappa^2H(y,x)}{4(1-C_2^2)(1-\mu)(x-y)^2}\left(\frac{dx^2}{G(x)}-\frac{dy^2}{G(y)}\right),
\end{eqnarray}
where 
\begin{eqnarray}
\omega_1&=&\frac{2\kappa ^2  C_1c(c-\mu)  (1+y)}{H(x,y)}  \biggl[-C_2^4(1-c) (c+\mu )  (1+x)^2\nonumber\\
&&+C_2^2 \biggl\{ -2(1-\mu)(2-\mu)-\mu^2x(y(1-x)+1+x)+\mu(2x^2+xy+x-y)\nonumber\\
&&+c\left(5-4\mu-\mu x^2(1+y)+(1-\mu)x(3-y)+y\right)
\biggr\}
+4(1-\mu)^4
\biggr], \nonumber\\
\omega_2&=&-\frac{2 \kappa ^2  C_1 C_2c  (c-\mu )(1+y)}{H(x,y)} 
\biggl[C_2^2\biggl\{\mu^2(xy+1)(1-x)+\mu(x^2-xy+5x+y-2)+4\nonumber\\
&&+c\left((1-2\mu+\mu y)x^2+((1-\mu)y-3\mu-1)x-y+\mu-4  \right)\biggr\}  -2(1-\mu)(\mu(x^2+2x-1)+2)
\biggr],\nonumber
\end{eqnarray}
and the functions $H(x,y)$, $F(x,y)$, and $J(x,y)$ are written, respectively,  as
\begin{eqnarray}
H(x,y)&:=&C_2^4 \biggl[-c^3 (x+1)^2 (1+y)-c^2 (x+1)^2 (1+y) (\mu(1+y)-1)\notag \\
&&-c \mu  (x+1)^2 (1+y) ((\mu -1) y-1)+\mu ^2 (x+1)^2 y (1+y)\biggr]\nonumber\\
&+&C_2^2 \bigg[c^2 (1+y) \{4 \mu +x (\mu  ((x-1) y+x+3)+y-3)-y-5\}\nonumber\\
&&+c \biggl \{-13 \mu +(\mu -1) \mu  x^2 (1+y)^2+2 x \left(\mu  (\mu +y (2 \mu +3 \mu  y-y+2)-1)+2\right)\nonumber\\
&&+\mu  (5 \mu +y (\mu  (y+2)+3 y-2))+8 y+12\biggr\}\nonumber\\
&&-\mu ^2 \bigl\{(x (x+3)+4) y^2+2 x (x+5) y+(x-1) x-4 y+8\bigr\}\nonumber\\
&&-\mu ^3 (x-1) (y-1)^2-4 \mu  (x+2 y-3)-8\biggr]\notag\\
&&-4 c (\mu -1)^2 (1+y)+4 (\mu -1)^2 (\mu  (y-1)+2),\nonumber
\end{eqnarray}
\begin{eqnarray}
\frac{F(x,y)}{\kappa^2}&:=&-2C_2^4\biggl[ (1-c)(1+x)^3(c+\mu)(c+\mu x) (1-y^2)(1+\mu y) \biggr]                      \notag \\ 
&+&C_2\biggl[ c^2 \biggl\{(-5 + x (-3 + y) - y) (1 + y) + (1 + y) (4 + x (3 + x + (x-1) y)) \mu\biggr\}\notag\\
&&+c \biggl\{4 (3 + x + 2 y) + (-13 - 2 x (y-1)^2 - 2 y + 3 y^2 - x^2 (1 + y)^2) \mu \notag \\
&&+ (5 + 2 y + y^2 + x^2 (1 + y)^2 + x (2 + 4 y + 6 y^2)) \mu^2\biggr\}\notag \\
&&- (x-1) (y-1)^2 \mu^3+(-8 - (x-1) x + 4 y - 2 x (5 + x) y - (4 + x (3 + x)) y^2) \mu^2\notag \\
&&-4 (-3 + x + 2 y) \mu-8\biggr]\notag\\
&&+8 (-1 + y^2) (1-\mu)^2 (2 - c (1 + x) + (x-1) \mu) (1 + y \mu),
\end{eqnarray}
\begin{eqnarray}
\frac{J(x,y)}{\kappa^2}&=& -2C_2^3 (-1 + c) (1 + x) (x - y) (1 + y) (c + \mu) \biggl[-4 + \biggl\{2 + x^2 (-1 + y)\notag \\
&& - 5 y - y^2 + x (-5 + y^2)\biggr\} \mu + \biggl\{-1 + y + x^2 (-1 + y) y - x (-1 + 4 y + y^2)\biggr\} \mu^2 \biggr] \notag\\
&+&4C_2(x - y) (1 - \mu) \biggl[2 (c (1 + x) (1 + y) - 2 (x + y))\notag \\
&& -  \biggl\{2 + 4 x^2 - 2 y + 4 y^2 + c (1 + x) (1 + x (-3 + y) - 3 y) (1 + y)+ 2 x (-1 + 5 y)\biggr\} \mu \notag\\
&&+  \biggl\{-1 - 2 y + 3 y^2 + x (-2 + 4 y - 6 y^2) - x^2 (-3 + 6 y + y^2)\notag  \\
&&+ c (1 + x) (1 + y) (1 - y + x (-1 + 3 y))\biggr\} \mu^2 \notag \\
&&-(1-x) (-1 + x (-1 + y) - y) (1 - y) \mu^3\biggr],
\end{eqnarray}

From Eq. (\ref{eq:condi3-2}), $z_1$ can be written in terms of the other parameters as
\begin{eqnarray}
z_1=\frac{z_2C_2^3+2z_4C_2^2+z_{32}C_2-2z_{43}}{C_2^3+2C_2^2}, \label{eq:z_1}
\end{eqnarray}
so that the substitution into Eq. (\ref{eq:condi4-2}) gives 
\begin{eqnarray}
\frac{z_{42}C_2^3-z_{32}C_2+2z_{43}}{z_{43}(C_2+2)^2}=1.\label{eq:m2b}
\end{eqnarray}
 It can be show that Eqs.~(\ref{eq:z_1}) and (\ref{eq:m2b}) have three different roots for $(C_2,z_1)$, which can be expressed in terms of only $(\mu,\kappa)$ as
\begin{eqnarray}
z_1&=&\kappa^2\mu,\\
C_2&=&-1,
\end{eqnarray}
and
\begin{eqnarray}
z_1&=&\frac{\kappa^2\mu(-9+7\mu\pm2\sqrt{9-8\mu^2})}{9(1-\mu)},\\
C_2&=&\frac{z_{42}+5z_{43}\pm\sqrt{z_2^2+z_3^2+34z_2z_3+36z_{42}z_{43}}}{2z_{32}}=\frac{3-2\mu\pm\sqrt{9-8\mu^2}}{2}.
\end{eqnarray}
Because of our assumption $z_1<z_2=-\kappa^2\mu$, we must take 
\begin{eqnarray}
z_1&=&\frac{\kappa^2\mu(-9+7\mu-2\sqrt{9-8\mu^2})}{9(1-\mu)}, \\
C_2&=&\frac{3-2\mu-\sqrt{9-8\mu^2}}{2\mu}.
\end{eqnarray}

\subsubsection{Asymptotic charges}

The mass and two angular momenta are given, respectively,  by
 \begin{eqnarray}
M=\frac{3\pi}{4}z_{31},\quad J_1=-\pi\sqrt{\frac{z_{21}z_{31}z_{41}(1-C_2^2)}{2}},\quad J_2=0.
\end{eqnarray}
The dimensionless angular momenta and horizon area are given by  
\begin{eqnarray}
&&j:=\sqrt{\frac{27\pi}{32}}\frac{J_1}{(GM)^{\frac{3}{2}}}=\sqrt{\frac{z_{21}z_{41}(1-C_2^2)}{z_{31}^2}},\quad j_2:=\sqrt{\frac{27\pi}{32}}\frac{J_2}{(GM)^{\frac{3}{2}}}=0,\\
&&  a_h:= \sqrt{\frac{27}{16^2\pi}}\frac{A_h}{(GM)^{\frac{3}{2}}}=\frac{2\sqrt{2}|z_{21}C_2^2+z_{32}|\sqrt{z_{32}z_{41}}}{z_{31}z_{42}\sqrt{(1-C_2^2)}}.
\end{eqnarray}

\subsubsection{Phase diagram}
From FIG.1, we see that the dimensionless angular momentum $j$ is a monotonically increasing function of $\mu$ and  $j\to\frac{1}{2\sqrt{2}}$ at $\mu\to 0$, and $j\to1$ at $\mu\to 1$. 
Figure 2  illustrates the relation between $j$ and the dimensionless horizon area $a_h$, where 
the black and blue curves correspond to the Myers-Perry (MP) black hole with a single angular momentum and the Emparan-Reall (ER) black ring, respectively, and the red curve corresponds to the black lens  with a single augular momentum.
The ER (thin) black ring has no upper bound for $j$, whereas the black lens has the upper bound $j=1$ (at $\mu=1$).
The MP black hole has a zero lower bound for the angular momentum $j$, whereas  the black lens has the non-zero lower bound $j=1/2\sqrt{2}$ (at $\mu=0$)  as the MP black hole does.
It turns out from these graphs that the dimensionless horizon area of the black lens is always larger than that of MP black hole (within the range of $1/2\sqrt{2}<j<1$) and $a_h$ vanishes at $j\to 1$ ($\mu\to 1$), which corresponds to the limit to the singular extreme MP black hole, and take the finite  value $2\sqrt{3}$ at $j\to 1/2\sqrt{2}$ ($\mu\to 0$), which is a singular solution without a horizon. 
Moreover, it can be seen that there indeed exists the parameter region such that $a_h$ of the black lens can exceed to those of the other three solutions, the black hole and thin/fat  black rings.


\begin{figure}[h]
 \centering
\includegraphics[width=10cm]{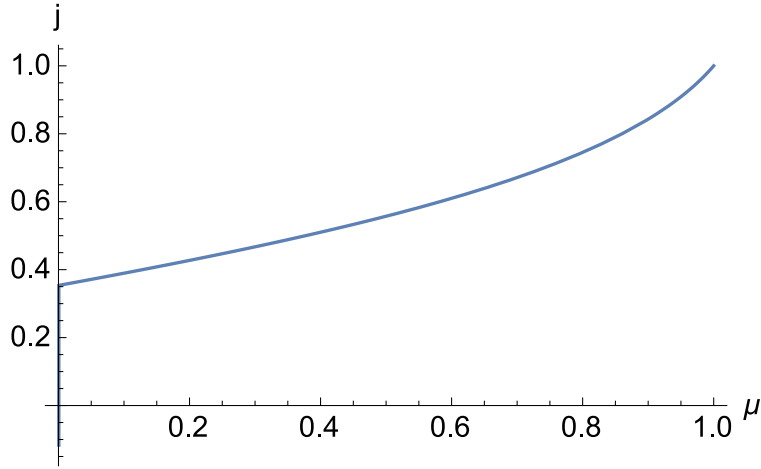}
\caption{Relation between $\mu$ and $j$. }
\end{figure}

\begin{figure}[h]
 \centering
\includegraphics[width=12cm]{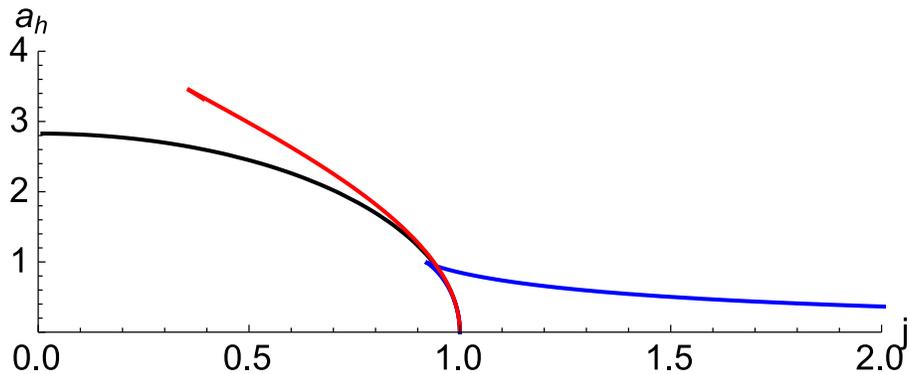}
\caption{The curves $a_h$ vs. $j$ for the five-dimensional Myers-Perry black hole, the Emparan-Reall black ring and the rotating black lens. 
The black curve and blue curve correspond to the MP black hole and the ER black ring, respectively and the red curve corresponds to the black lens. }
\end{figure}

\subsubsection{CTCs and curvature singularities}
We wish to require absence of CTCs  in the domain of outer communication.
The necessary and sufficient conditions to ensure that CTCs do not exist in the domain of outer communication is such that the following two-dimensional matrix always becomes nonnegative in the region:
\begin{eqnarray}
g_2=\left(
    \begin{array}{cc}
    g_{\phi_1'\phi_1'} & g_{\phi_1'\phi_2'} \\
      g_{\phi_1'\phi_2'}  &g_{\phi_2'\phi_2'}  
    \end{array}
  \right),
 \end{eqnarray}
namely, CTCs do not exist if and only if 
\begin{eqnarray}
{\rm det}\ g_2\ge 0,\quad {\rm Tr}\ g_2\ge 0. 
\end{eqnarray}
We have numerically studied the positivity for various values of $\mu$ and the normalized $\kappa$ (as $\kappa=1$) in the $(\rho,z)$-plane. 
In FIG.7, the point $(\rho,z)=(0,1)$ and the interval on the $z$-axis $\{(\rho,z)\ |\ \rho=0,  -n/10\le z \le n/10\}$ represent the turning point ($(\rho,z)=(0,z_4)$) and the Killing horizon, respectively, for $\mu=n/10\ (n=1,4,7,9)$.
In each figure, the white region represents the CTC region, which always appears around the turning point $(\rho,z)=(0,z_4)$.  
We have confirmed that regardless of the values of $\mu$, the white region which surrounds the point $(\rho,z)=(0,z_4)$ exists outside the horizon.  
As a consequence, it can be seen that the existence of CTCs outside the horizon seems to be unavoidable.

Next, let us see if curvature singularities exist inside or outside the horizon. 
For this purpose, we consider where one of scalar invariants, for instance,  Kretschmann invariant  $R_{\mu\nu\rho\sigma}R^{\mu\nu\rho\sigma},   $ diverges in the $(\rho,z)$-plane.
One can find from a direct computation  that it diverges at the points $(x,y)$ satisfying $H(y,x)=0$, which is denoted by the red curve in each figure of FIG.7.   
It can be seen from these figures that $H(y(\rho,z),x(\rho,z))=0$ holds just on the spherical boundary of the CTC region around the point $(\rho,z)=(0,z_4)$. 
Therefore, it can be concluded that the black lens with a single angular momentum unavoidably has curvature singularities outside the horizon and on the spherical boundary of the CTC region.

\begin{figure}[h]
 \begin{tabular}{cc}

 \begin{minipage}[t]{0.4\hsize}
 \centering
\includegraphics[width=7cm]{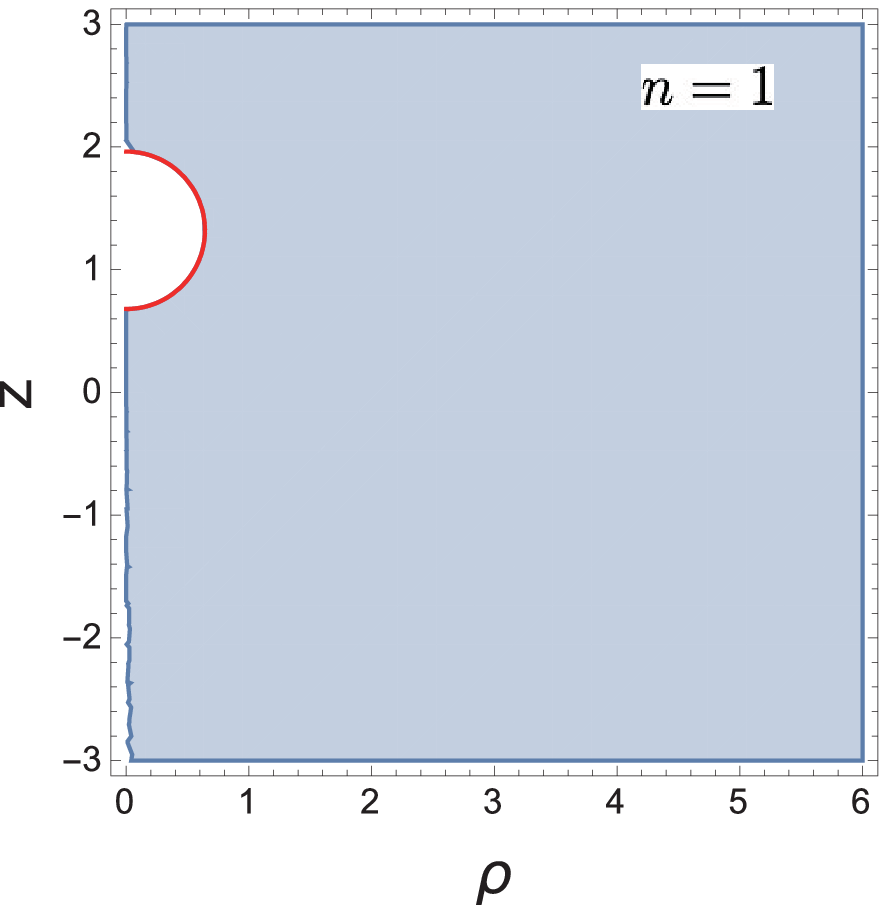}
 \end{minipage} &\ \ \ \ \ \ \ \ \ \ \ \ \ \ \ \ \ 

\begin{minipage}[t]{0.4\hsize}
 \centering
\includegraphics[width=7cm]{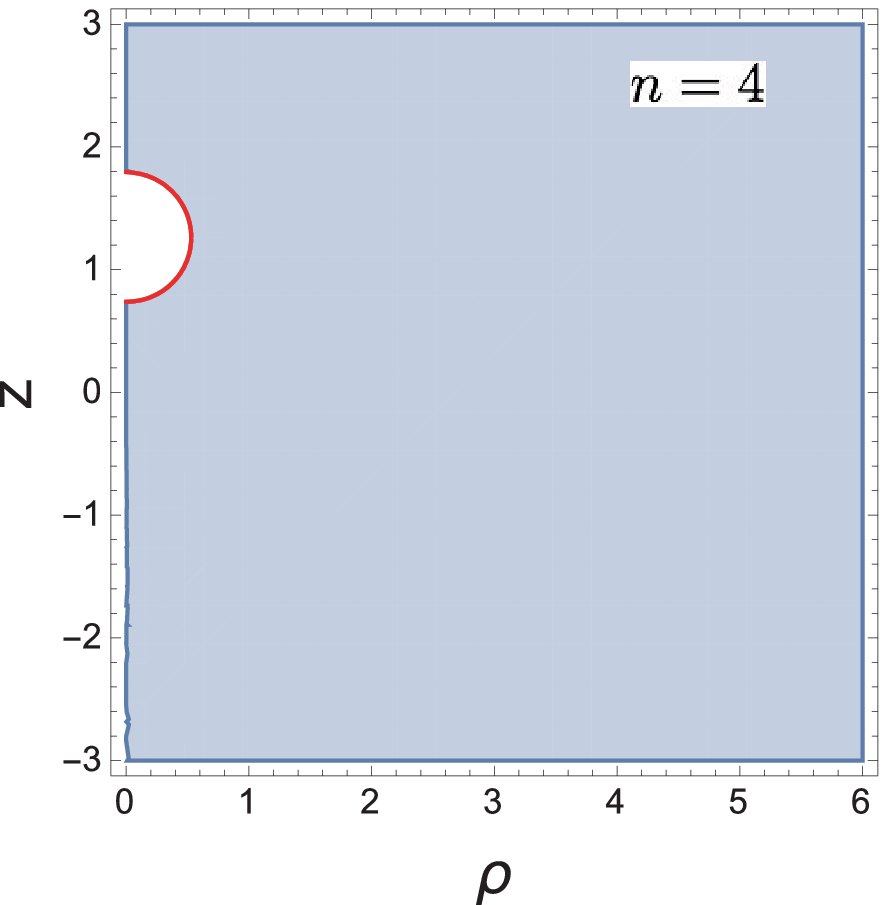}
 \end{minipage} \\&\\

\begin{minipage}[t]{0.4\hsize}
 \centering
\includegraphics[width=7cm]{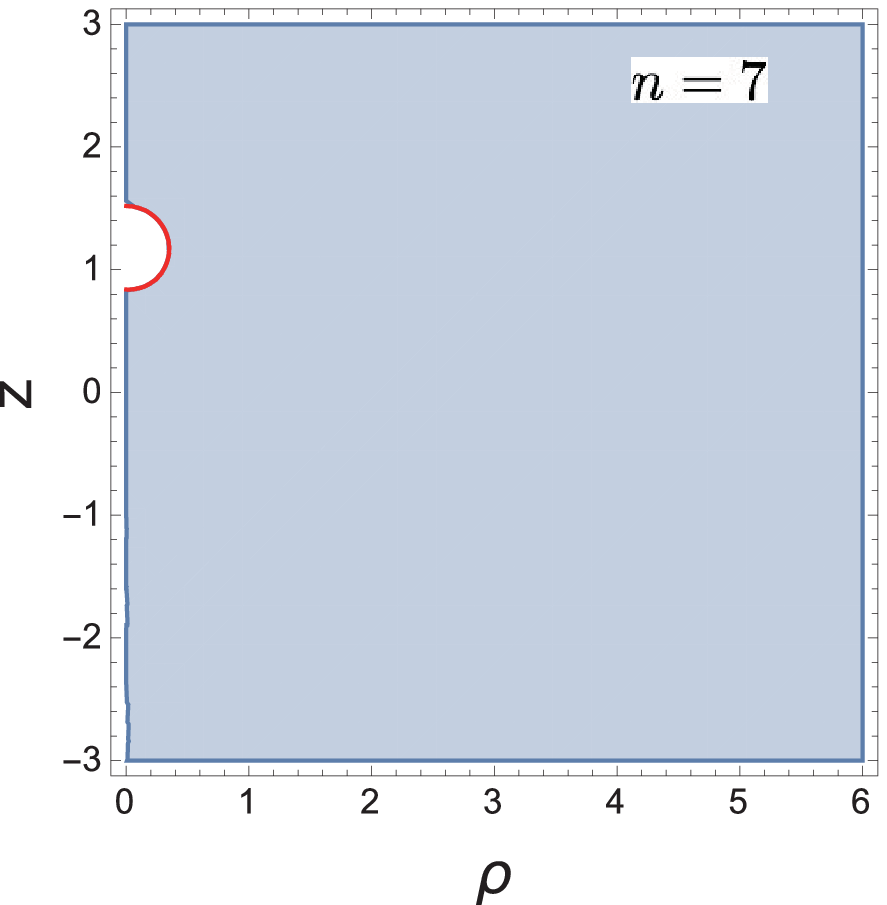}
 \end{minipage} &\ \ \ \ \ \ \ \ \ \ \ \ \ \ \ \ \

 \begin{minipage}[t]{0.4\hsize}
 \centering
\includegraphics[width=7cm]{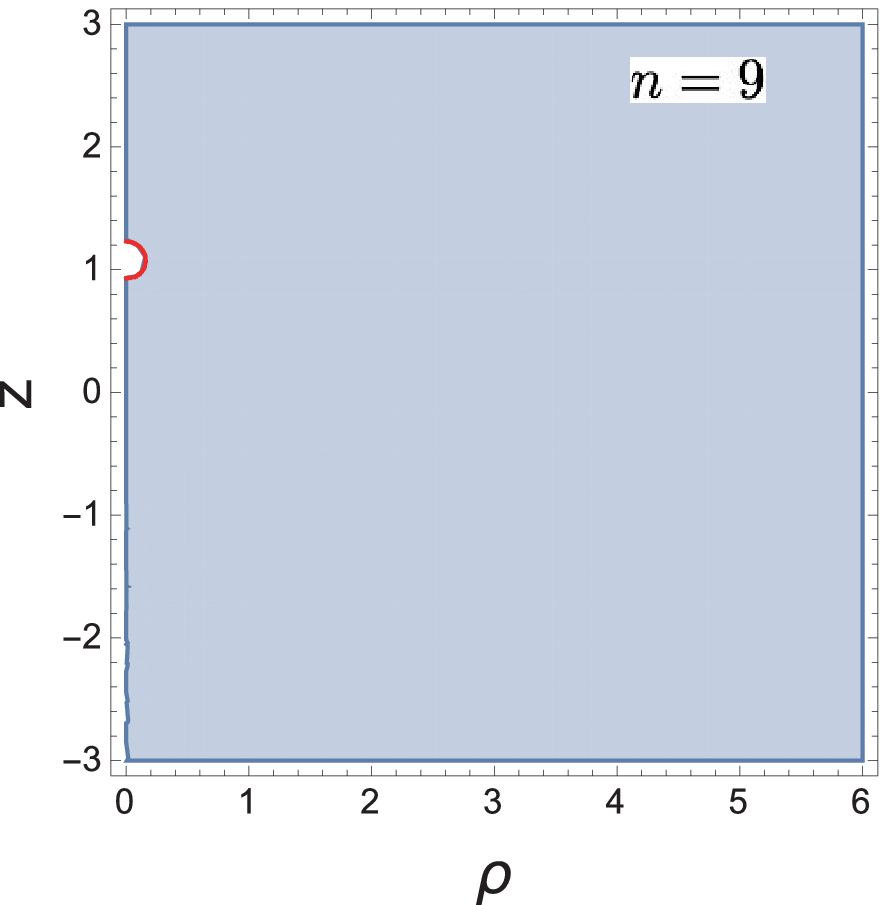}
 \end{minipage}

\end{tabular}
\label{fig:ctc}

\caption{In each figure, the white region represents the CTC region for $\mu=n/10\ (n=1,4,7,9)$. 
There exist curvature singularities on the red curve (spherical surface) which is the boundary of the CTC region. 
The point $(\rho,z)=(0,1)$ and the interval $(\rho=0,-n/10\le z \le n/10)$ represents the nut and the Killing horizon, respectively. }
\end{figure}

\newpage
\subsection{Static solution}
Finally, we consider the static limit of the four-soliton solution $z_1\to z_2, C_3\to0$, in which case it follows from Eq.~(\ref{eq:condi1}) that $C_1\to 0$ holds.  
In terms of the $C$-metric, the metric takes the following simple form
\begin{eqnarray}
ds^2&=&-\frac{1+\mu y}{1+\mu x}dt^2+\frac{2\kappa^2(1+\mu x)[C_4^2(1-x^2)(1+\mu y)^2+(1-\mu)^2(y^2-1)]}{(x-y)^2[(1-\mu)^2-C_4^2(1+\mu x)(1+\mu y)]}d\phi_1^2\notag\\
&&+\frac{2\kappa^2(1+\mu x)[C_4^2(y^2-1)(1+\mu x)^2-(1-\mu)^2(1-x^2)]}{(x-y)^2[(1-\mu)^2-C_4^2(1+\mu x)(1+\mu y)]}d\phi_2^2\notag\\
&&-4\frac{\kappa^2C_4(1-\mu )(1+\mu x)[x+y+\mu(1+xy)]}{(x-y)[(1-\mu)^2-C_4^2(1+\mu x)(1+\mu y)]}d\phi_1d\phi_2\notag\\
&&+\frac{2\kappa^2(1+\mu x)[(1-\mu)^2-C_4^2(1+\mu x)(1+\mu y)]}{(1-C_4^2)(1-\mu)(x-y)^2}\left(\frac{dx^2}{G(x)}-\frac{dy^2}{G(y)}\right),
\end{eqnarray}
where it should be noted that  $C_2$ automatically disappears.
In the limit,  the condition (\ref{eq:condi-C_4}) is automatically satisfied, and  two equations (\ref{eq:condi-lens}) and (\ref{eq:condi-conical}) are simplified, respectively, as
\begin{eqnarray}
&&\frac{2\mu C_4}{(1-\mu)-C_4^2(1+\mu)}=2,\label{eq:condi-lens-static}\\
&&\frac{[1-\mu-C_4^2(1+\mu)]^2}{(1-C_4^2)^2(1-\mu^2)}=1.\label{eq:condi-conical-static}
\end{eqnarray}
From Eq.~(\ref{eq:condi-lens-static}), $\mu$ can be written as
\begin{eqnarray}
\mu=\frac{-C_4^2}{C_4^2+C_4-2},
\end{eqnarray}
and then Eq.(\ref{eq:condi-conical-static}) becomes 
\begin{eqnarray}
\frac{C_4}{2C_4^2+5C_4+2}=1,
\end{eqnarray}
which gives $C_4=-1$. 
This result contradicts with Eq.(\ref{eq:condi-conical-static}). 
Therefore, this implies that when Eq.~(\ref{eq:condi-lens-static}) holds, Eq.(\ref{eq:condi-conical-static}) cannot be satisfied. 
We can interpret physically  that the static black lens needs conical singularities on the rod $[z_3,z_4]$ to support the horizon against gravitational attraction.
It may be of interest to investigate such a static black lens even though it has conical singularities. 
From Eq.~(\ref{eq:condi-lens-static}), $C_4$ can be solved as
\begin{eqnarray}
C_4=C_{4\pm}:=\frac{-\mu\pm\sqrt{4-3\mu^2}}{2(1+\mu)}.
\end{eqnarray}
Hence, the static black lens solution has two branches according to the choice of $C_4$.
It can be shown numerically that for $C_4=C_{4+}$, CTCs do not exist, whereas for  $C_4=C_{4-}$,  there always exists CTC region which  surrounds the point $(\rho,z)=(0,z_4)$. 
It can be shown that after performing  the global rotation, this  solution coincides with the static limit of the Chen-Teo's solution~\cite{Chen:2008fa} in which the parameters $a$, $c$, $n$ and the angular coordinates ($\psi$, $\phi$) are replaced, respectively,  with $C_4$, $\mu$,  $-2$, and ($-\phi_1'$,$\phi_2'$).

\newpage
\section{Summary}
\label{sec:summary}
Using the ISM for the five-dimensional vacuum Einstein equations and starting from the same seed as for the black ring construction, we have obtained the four-soliton solution,  which includes an asymptotically flat,  stationary and biaxisymmetric black hole with the horizon topology of $L(2,1)=S^3/{\mathbb Z}_2$.  
The solution has nine parameters $(C_i,z_i,k)$ which obey to five constraint equations imposed from the physical requirements. 
Therefore,  except for the gauge freedom  $z\to z+\alpha$, the remaining degrees of freedom  reduce to three,  which physically correspond to the mass and two angular momenta. 
 The rod diagram of the obtained solution is the same as that of the Kunduri-Lucietti's supersymmetric black lens except for the horizon rod rather than that of Chen-Teo's solution. 
We wish to emphasize that the rod diagram is different from that of Chen-Teo's solution in whether the orientation of the finite rod between the horizon and the nut is $(0,2,1)$ or $(0,2,-1)$, where Chen and Teo considered the former, whereas we have chosen the latter. 
We have also showed that there is the parameter region such that on the axes of symmetry  there exist no curvature singularities, no conical singularities, and  no orbifold singularities. 

\medskip
Since the metric of the four-soliton solution takes a considerably complicated form even in the $C$-metric representation,  we have analyzed, in particular, the static case and the case of a single angular momentum, which corresponds to a two-soliton solution and a three-soliton solution, respectively. 
We have discussed the phase diagram of the black lens with a single angular momentum  in comparison with those of the MP black hole with a single angular momentum and the ER black ring. 
In contrast to the ER (thin) black ring,  the angular momentum for the black lens has the upper bound, and 
in contrast to the MP black hole with a single angular momentum, it has a certain non-zero lower bound.
We have shown that there exists the parameter region such that the four different solutions, the black hole, thin/fat black ring and the black lens, exist for the same asymptotic charges (mass and an angular momentum).
Unfortunately, we have found that for the case of single angular momentum, the existence of CTCs and curvature singularities around the nut cannot be avoidable. 
The static black lens solution whose horizon topology is $L(2,1)=S^3/{\mathbb Z}_2$ can be obtained by simply setting $z_1= z_2$ and $C_3=0$.  
We have shown that the solution always has conical singularities between the horizon and the nut, but there are two possibilities for this static solution. 
 One has an unavoidable CTC region which includes the nut, while the other does not admit  any CTCs  in the domain of outer communication.

\medskip
The four-soliton solution with $C_3\not=0$, which we have obtained in Sec.\ref{sec:solution}, has much more complicated metric  than the three-soliton solution with $C_3=0$ even if we use the $C$-metric representation. 
This makes it difficult that we even numerically check the existence of CTCs and regularity for the far region from the rods and horizon through the whole parameter region.  
We have numerically found that for the 4-soliton solution, there also exist the parameter region where there are neither conical singularities nor curvature singularities on the axis and horizon, but it seems that CTCs inevitably appear around the nut for several special sets of the parameters.  
To see if there are CTCs for the whole parameter region is our remaining future study.

\medskip
From our results in this paper, we cannot conclude immediately that a supersymmety plays an essential role in the existence of a regular black lens without naked CTCs simply because such a black lens has been found only within a class of sympersymmetric solutions so far. 
As for the supersymmetric solutions~\cite{Kunduri:2014kja,Tomizawa:2016kjh}, the existence of magnetic fluxes rather than an electric charge seems to be essential to support the horizon of the black lens. 
Therefore, it may be also possible that a non-BPS black lens with a magnetic flux exists in five-dimensional minimal supergravity. 
Moreover, it may be an interesting issue whether a vacuum solution of the more general horizon topology $L(p,q)$ ($p,q$: comprime integers) under the same symmetry assumptions, since it was shown in Ref.~\cite{Breunholder:2017ubu} that such a general black lens cannot exist, at least, within a class of  asymptotically flat, stationary and bi-axisymmetric supersymmetric solutions in five-dimensional minimal supergravity. 
These issues deserve further study.

\acknowledgments
This work was supported by Grant-in-Aid for Scientific Research (C) (No.~17K05452) from Japan Society for the Promotion of Science (S.T.).


\newpage


\end{document}